\documentclass[preprint,12pt]{elsarticle}
\usepackage{amssymb}

\usepackage{xcolor}
\usepackage{booktabs} 
\usepackage{ulem}
\usepackage{subfigure}
\usepackage{dsfont}
\usepackage[algo2e,inoutnumbered,linesnumbered,algoruled,vlined]{algorithm2e}
\usepackage{graphicx}
\usepackage{bbm}
\usepackage{multirow}
\usepackage{amsmath}
\usepackage{setspace}
\usepackage[colorinlistoftodos,textsize=tiny]{todonotes}

\graphicspath{{images/}}

\definecolor{mydarkblue}{rgb}{0,0.08,0.45}

\newcommand{\website}{https://nilm.telecom-paristech.fr/shed/}

\DeclareMathOperator{\volt}{\mathbf{u}}
\DeclareMathOperator{\curr}{\mathbf{i}}
\DeclareMathOperator{\currDFT}{\mathbf{I}}
\DeclareMathOperator{\pwr}{\mathbf{p}}
\DeclareMathOperator{\sig}{{\mathbf{s}}}
\DeclareMathOperator{\act}{{\mathbf{a}}}
\journal{Energy and Building}

\begin{document}
\title{A Generative Model for Non-Intrusive Load Monitoring in Commercial Buildings}

\date{\today}

\author[tel,si]{Simon Henriet}

\author[tel]{Umut \c Sim\c sekli}

\author[si]{Benoit Fuentes}

\author[tel]{Ga\"el Richard}
\address[tel]{LTCI, T\'el\'ecom Paristech, Universit\'e Paris Saclay, Paris, France}
\address[si]{Smart Impulse, Paris, France}


\begin{abstract}
In the recent years, there has been an increasing academic and industrial interest for analyzing the electrical consumption of commercial buildings. Whilst having similarities with the Non Intrusive Load Monitoring (NILM) tasks for residential buildings, the nature of the signals that are collected from large commercial buildings introduces additional difficulties to the NILM research causing existing NILM approaches to fail. On the other hand, the amount of publicly available datasets collected from commercial buildings is very limited, which makes the NILM research even more challenging for this type of large buildings. In this study, we aim at addressing these issues. We first present an extensive statistical analysis of both commercial and residential measurements from public and private datasets and show important differences. Secondly, we develop an algorithm for generating synthetic current waveforms. We then demonstrate using real measurement and quantitative metrics that both our device model and our simulations are realistic and can be used to evaluate NILM algorithms. Finally, to encourage research on commercial buildings we release a synthesized dataset.

\end{abstract}


\maketitle

\section{Introduction}
With the increasing awareness about the problem of climate change and the increasing level of energy consumption, a need for energy efficiency has emerged. At the Paris conference of the parties (COP21) \cite{protocol2015report}, many countries have recognized energy efficiency as the basis of energy transition.

An important step towards energy efficiency is based on reducing the energy consumption in residential and commercial buildings. To this end, one needs to measure and analyze the power consumption profiles of the devices that are installed in the buildings. In this context, depending on the particular application, one can be interested in either the estimation of the total energy consumed by devices in a certain period, the operational schedule of a particular device, the power consumption profile at a certain time-step or the estimation of electrical reliability.

There are two main research directions for electrical load monitoring: (i) full sub-metering and (ii) non-intrusive load monitoring (NILM). The former requires installing a sub-meter on each electrical device plugged into the network. While being accurate, this approach has important financial and computational limitations since it requires an excessive amount of measurement devices. On the contrary, the latter, the main subject of this study, involves only one sensor per building, installed at the entrance of the electrical network and therefore has a much less demanding data collection process. However, since the measured signals contain information coming from all the devices, NILM requires accurate energy disaggregation algorithms for estimating the electrical consumption of each device.

The majority of the current NILM literature has been focused on residential buildings using low frequency ($< 50$ or $60$ Hz) power data obtained by smart meters. Recently, there has been an increasing academic and industrial interest in applying NILM to commercial buildings. These buildings include large offices, warehouses, retails or shopping malls, and as also pointed out in \cite{batra2014comparison}, have fundamentally different characteristics than those of residential buildings. On top of this, the use of high frequency measures (such as current waveforms) has been enabled by the development of low cost meters \cite{quitana2017demo} and can significantly change the classic power based NILM approaches.

An important limitation for developing disaggregation algorithms for commercial buildings is the lack of publicly available datasets that contain detailed measurements of individual devices collected from commercial buildings. Unfortunately, collecting such data turns out to be a very challenging and expensive task since it requires installing sensors on each device in a large building, and the quality of these measurements is difficult to be maintained during a long period. To the best of our knowledge, there is only one public dataset that is collected from a commercial building, namely the COMBED dataset \cite{batra2014comparison}. This dataset contains the power consumption measurements of two buildings (an academic and a library blocks) and is sampled at $1/30$ Hz. Even though it is a first step towards energy disaggregation in commercial buildings, the dataset does not include high frequency data (current or voltage) and the equipments are not fully sub-metered.
As a consequence of the lack of data, developing supervised machine learning algorithms for NILM becomes even more challenging. Unsupervised algorithms require less data to be developed but still need data of good quality for evaluation purposes.

In this paper, we propose a comprehensive framework for energy disaggregation in commercial buildings. We aim at circumventing the issues caused by the lack of knowledge and data available. 
We first perform a statistical analysis on public residential datasets and compare them to a private dataset that is collected from real commercial buildings in France, in order to have a better understanding of the differences between the two kinds of buildings.
In the light of the results obtained from our analyses and by making use of both the publicly available datasets and the private dataset, we develop a synthetic data generation algorithm that is able to produce realistic high frequency current waveforms. We then conduct various experiments for evaluating the quality of both our device models and building simulations.
To finally foster the NILM research for commercial buildings, we release a synthetic dataset, called SHED\footnote{\website}, that is generated by our algorithm.

\section{Related work}
NILM formalism for residential building has first been introduced in two papers by Hart and Sultanem \cite{hart1992,sultanem1991using}. Hart studied electrical devices from a low frequency power consumption point of view and classified them as: (i) on/off or constant device, (ii) multi-state and (iii) continuously varying. Sultanem worked on high frequency current measurements and introduced the notion of harmonic content to cluster devices into category such as: (i) resistive, (ii) motor-driven, (iii) electronically-fed or (iv) fluorescent lighting.

Following these first papers, many NILM algorithms have been developed for residential buildings. They all share the same major features: (i) a power disaggregation step and (ii) a load classification step. In this context, they can be classified into two categories. The first category, `event-based disaggregation' techniques, aims at detecting certain events (e.g.\ on/off transition, change of state) and then assign the detected events to different electrical devices \cite{hart1992,laughman2003power,girmay2016simple,barsim2015toward}. It relies on two assumptions: (i) `one at time', at most one device changes of state at each instant, and (ii) `constant load', the consumption remains constant between two events. The second category, `source separation' techniques, assumes that the measured signal is a \textit{mixture} of unknown \textit{source} signals that correspond to different devices. The goal is then to recover the source signals from the mixture signal. Most of the proposed methods use low frequency power measurements \cite{kolter2012approximate,kolter2010energy,rahimpour2017non} whereas Lange and Berges are using source separation on high frequency current measures \cite{lange2016bolt}.

NILM for commercial buildings started with Norford and Leeb's work \cite{norford1996non}. They identified three main challenges for tackling commercial buildings: (i) load detection; due to the recurrence of overlapping events (switching on or off) (ii) load estimation; due to variations in load for several devices and (iii) load identification; due to similarity in low frequency features for different devices. Batra et al. also pointed out that the hypotheses made by existing `event-based' NILM approaches do not hold in this context and showed that NILM algorithms developed for residential buildings fail when applied to commercial buildings \cite{batra2014comparison}. To overcome low frequency data limitations, Lee et al. used current harmonic content to separate variable speed drives from aggregate data in commercial buildings \cite{lee2005estimation}. As underlined by Zeifman and Roth \cite{zeifman2011nonintrusive}, there is a large agreement that raising the sampling frequency also increases the probability of getting accurate estimation of individual consumption \cite{dong2014fundamental}. Since in commercial buildings we are facing a larger number of devices, high frequency measurements can also help in the disaggregation task.

Substantial efforts have been made to model electrical devices consumption and simulate datasets in order to evaluate NILM algorithms. In \cite{fischer2015model}, they use "on/off" models with a probability of a device to be switched on depending on the time of the day. Other approaches \cite{buneeva2017ambal,barker2013empirical} define more complicated models that can take into account uniform randomness during operation time, multi-state devices or exponentially decaying load curves. Even though these models are efficient for electrical devices in residential buildings, they lack of complexity to be used  in commercial settings which contains smoothly varying devices and need high frequency measurements.

It is worth mentioning that high frequency current measures have been studied in several papers \cite{sultanem1991using,lee2004exploration,liang2010load}. Lam et al. used high frequency current/voltage trajectories to classify electrical devices \cite{lam2007novel}. Public datasets of high frequency current measurements of residential equipments has also been released \cite{gao2014plaid,picon2016cooll}.

Finally, Liang et al. \cite{liang2010load} developed a simulator for high frequency current measures but without considering long term modeling of current dynamics.

\section{Statistical Analysis of Residential and Commercial NILM Datasets}
  \label{sec_analysis}
\subsection{Datasets}
In most commercial or residential buildings, the electric power is delivered as alternating current (AC) (sinusoidal voltage) and distributed with 1, 2 or 3 phase lines, corresponding to fix voltage phases difference. The different quantities that can be measured by the sensors are energy per period (kWh), instantaneous or average real power in watt (W) or current and voltage in ampere (A) and volt (V). These quantities are related to the notion of sampling frequency. A common definition in the literature is to consider as high frequency (HF) a measurement occurring multiple times within an electrical period (defined by the fundamental frequency of the voltage) and as low frequency (LF) a measurement that occurs at a lower frequency than the fundamental. HF measurements generally correspond to current and voltage whereas LF measurements correspond to power or energy.

\begin{table}[ht]
  \centering
  \caption{\label{<datasets>} The public and private datasets used in this study.}
  \begin{tabular}{|l|c|c|c|c|c|}
      \hline
      Name & Data & Buildings & Phases & Frequency  & Type \\
      \hline
      BLUED \cite{filip2011blued}& current & 1 & 2 & 12 kHz & residential \\
      UK-DALE \cite{kelly2015uk-dale} & current & 1 & 1 & 16 kHz & residential \\
      REDD \cite{kolter2011redd}& current & 2 &  2/1 & 16.5 kHz & residential \\
      SIHF [private]& current & 7 & 3 & kHz & commercial \\
      REDD \cite{kolter2011redd}& power & 6 & 2 & 1 Hz & residential \\
      ECO \cite{beckel2014eco}& power & 6 & 1 & 1 Hz & residential \\
      IAWE \cite{batra2013s}& power & 1 & 1 & 1 Hz & residential \\
      UK-DALE \cite{kelly2015uk-dale}& power & 5 & 1 & 1/6 Hz & residential \\
      REFIT \cite{murray2017electrical}& power & 20 & 1 & 1/8 Hz & residential \\
      RAE \cite{makonin2017rae}& power & 1 & 2 & 1/15 Hz & residential \\
      COMBED \cite{batra2014comparison} & power & 1 & 1 & 1/30 Hz & commercial \\
      SILF [private] & power & 7 & 3 & 1/30 Hz & commercial \\
      \hline
  \end{tabular}
\end{table}

In the last decade, we have witnessed the release of multiple publicly available datasets of different quality and with different sampling strategies. In this section, our goal is to compare residential to commercial buildings from a statistical point of view at both high and low frequency (at least $1/30$ Hz). The public datasets used for this study range from low frequency \cite{batra2014comparison,kelly2015uk-dale,kolter2011redd,beckel2014eco,batra2013s,murray2017electrical,makonin2017rae} to high frequency sampling \cite{filip2011blued,kelly2015uk-dale,kolter2011redd} and correspond to measurements of individual houses (except for one which comes from an university building \cite{batra2014comparison}). From each dataset we have selected houses or buildings whose measurements last longer than a week. In addition to public data, two private datasets are used. It consists of both low frequency total power data (named SILF) and high frequency total current measurements (named SIHF) from 7 commercial buildings. All those datasets are shown in Table \ref{<datasets>}.

\subsection{Physical preliminaries}
Before getting to the statistical analysis, we shall introduce some notations and recall the relation between physical quantities. The digitized voltage and current waveforms are denoted: $\volt \left( n, t \right) \textrm{ and } \curr \left( n, t \right)$, where $t = 1, \dots, T$ is the voltage period index, $T$ denotes the total number of voltage periods, and $n$ is the sampling index within a voltage period. The number of samples within a period of the voltage sine wave is supposed to be constant and is noted $N$. This segmentation according to the voltage period enables us to have a matrix representation of both current and voltage. The mean active power (or mean power consumption or load curve) for a voltage period is then given by: 
\begin{equation} \label{eq:power_def}
\pwr(t) = \frac{1}{N}\sum_{n=1}^{N} \volt(n, t) \curr(n,t).
\end{equation}
It is possible to down-sample or aggregate this signal by averaging several consecutive periods (in order to have a sample every $30$ seconds for instance). For the sake of clarity, the same index $t$ is kept regardless of the sampling frequency. 

\subsection{Power measurements (low frequency)}
In order to discriminate residential buildings from commercial buildings, we are particularly interested in state change events, switching on/off events or continuous variations of electrical devices present in the building.  These events result in total current signal variations and therefore, due to equation \eqref{eq:power_def}, in a time-varying power consumption. In this section, we used all the power and current datasets presented in Table \ref{<datasets>}. Power values have been calculated according to \eqref{eq:power_def} for current datasets. Power time series exhibit a strong temporal structure, characterized by high first-order autocorrelation (0.92 and 0.99 in average for respectively residential and commercial buildings at $1/30$ Hz). This can be explained by the fact that, when a device is switched on it often remains active for several periods. This motivates us to study the power derivative rather than the power consumption: 
\begin{equation}
\pwr^\prime \left( t \right) = \pwr\left( t \right) - \pwr \left( t - 1 \right),
\end{equation} 
and to characterize its structure at different time scale. To enable the comparison between buildings, the power derivative is normalized so that the mean is zero and the standard deviation is one.

\begin{figure}[t]
  \centering
  \includegraphics[width=0.5\textwidth]{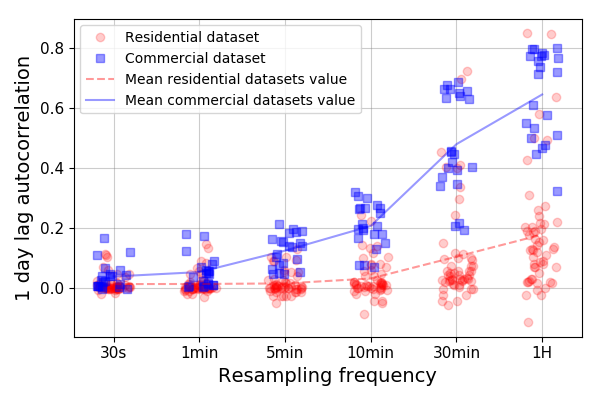}
  \caption{\label{<autocor>}Estimation of the 1 day lag autocorrelation for the power derivatives at different re-sampling frequencies for all the datasets (see Table \ref{<datasets>})}
\end{figure}

One important structure in time series is the seasonality. It is a weak assumption to state that the power consumption and thus its derivative can show daily seasonality due to the habits of the people and time-scheduled equipments. The serial autocorrelation with a lag of 1 day of the power derivative is presented in Figure \ref{<autocor>}. It first shows that the derivative of hourly aggregated power discriminates the two kinds of buildings, since the seasonal effect is higher for the commercial ones than for the residential ones ($0.65$ vs $0.18$ in average). This can be interpreted by the fact that the consumption patterns are more periodical in commercial buildings than in residential: (i) many equipments are programmed and have recurrent patterns, (ii) the average behavior of occupants is more recurrent than individual behaviors. Figure \ref{<autocor>} also shows that the seasonal effect is more intense at higher time scale.

\begin{figure}[t]
  \centering
  \includegraphics[width=0.5\textwidth]{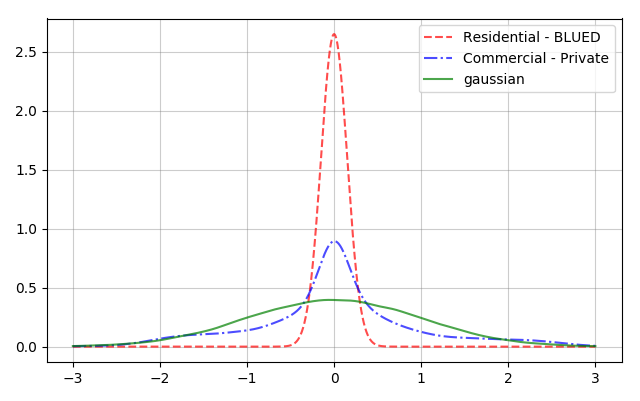}
  \caption{\label{<distribution>} Distribution of derivative power @ $1/30$Hz for all the datasets (see Table \ref{<datasets>})}
\end{figure}

At a $1/30$ Hz sampling frequency, the power derivative has almost no temporal structure (zero first-order autocorrelation) and can thus be studied as realizations of independent and identically distributed random variables.
It can be observed in Figure \ref{<distribution>} that the distribution of the power derivative for a residential building can be more peaky around zero and has a heavier tail than the one of a commercial building. Additionally, 3 statistics that accurately reflect the difference in distributions are presented: (i) the kurtosis, (ii) the entropy and (iii) the scale parameter of Laplace distribution.

\begin{figure*}[t]
  \centering
  \subfigure[Kurtosis]{\includegraphics[width=0.3\textwidth]{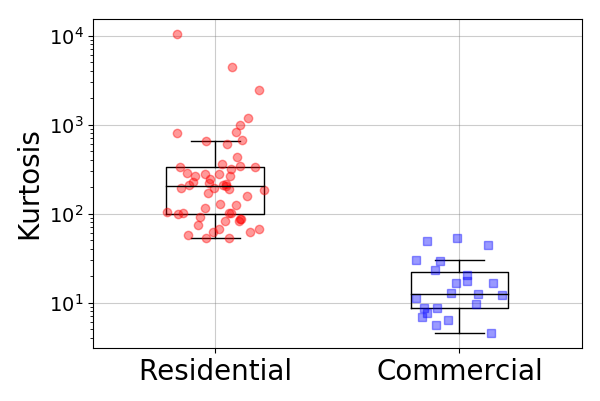}} 
  \subfigure[Entropy]{\includegraphics[width=0.3\textwidth]{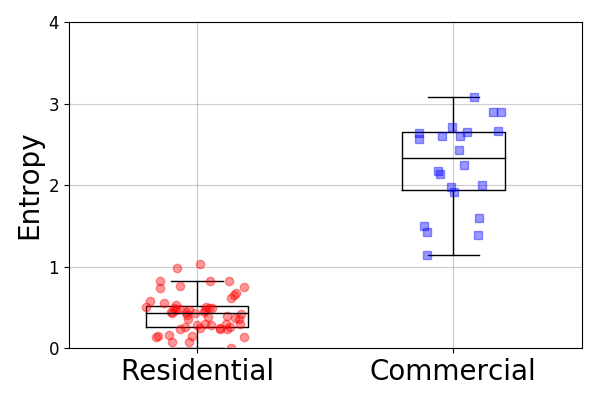}}
  \subfigure[Laplace scale parameter]{\includegraphics[width=0.3\textwidth]{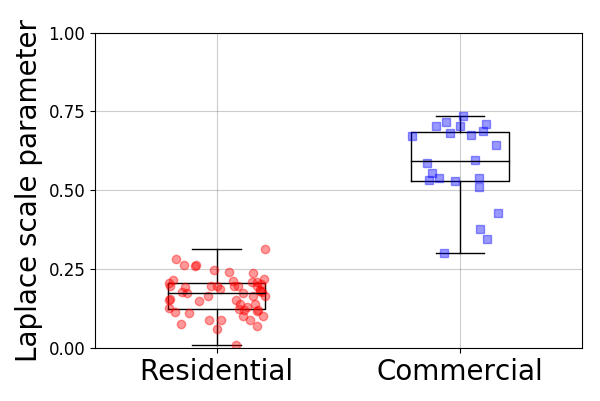}}
  \caption{\label{<low_stats>}Statistical analysis of power changes at a 1/30 Hz sampling frequency for all the datasets (see Table \ref{<datasets>})}
\end{figure*}

Firstly, the kurtosis is based on a scaled version of the fourth moment of a distribution:
\begin{align}
\text{Kurt}[X] = \frac{\mathds{E}\left[\left(X-\mathds{E}\left[ X \right] \right)^4\right]}{\mathds{E}\left[\left(X-\mathds{E}\left[ X \right] \right)^2\right]^2},
\end{align}
where $\mathds{E}$ is the mathematical expectation and $X$ a random variable.
It can be noted here that the kurtosis has often been used as a measure of impulsiveness: impulsive signals typically have a high kurtosis value \cite{liang2008statistical}.
Figure \ref{<low_stats>} shows a clear difference in kurtosis for the two types of building. 
On one hand, high kurtosis value for residential buildings can be explained by low number of devices and simple devices (ON/OFF or multi-state) which result in more impulsive power derivative signals. On the other hand, due to the central limit theorem, the more independent individual devices there are, the closer the random variable resulting from the sum is to a Gaussian. It explains why kurtosis values for commercial buildings are closer to the standard Gaussian kurtosis value ($3$) than kurtosis values for residential buildings.
It can however be observed that the kurtosis for commercial buildings remains high compared to the kurtosis of the standard Gaussian distribution, and this characteristic can still be used in NILM algorithms.

Secondly, entropy is defined as the average amount of information produced by a stochastic source of data. It is based on the logarithm of the probability density:
\begin{align}
\text{H}[X] = \mathds{E}\left[- \ln(P(X))  \right],
\end{align}
Figure \ref{<low_stats>} shows that entropy values are higher for commercial buildings. This results from the fact that commercial datasets contain more devices and thus more information, which is more complex to encode. This can also come from the fact that there are much more devices with varying power in commercial  buildings than in residential ones.

Finally, Laplace distribution is a high kurtosis distribution that has two parameters: a location ($\mu$) and a scale ($b$). The location parameter equals the mean of the distribution and is of less interest because it is $0$ for our normalized power derivatives. In order to compare the datasets, we estimate the scale parameter considering the distributions as Laplace and then compare the estimated parameters. A maximum likelihood estimator of the scale parameter is given by:
\begin{align}
\hat{b} = \frac{1}{N}\sum_{n=1}^{N}\lvert x_i-\mu \rvert,
\end{align}
As shown in Figure \ref{<low_stats>}, the estimated scale parameters are higher for commercial buildings. We can finally remark that these 3 criteria promote sparseness in the data.\footnote{For Laplace distributed random variable, entropy and the scale parameter are linked: $\text{H}[X]=\log(2\text{b}\text{e})$.}

\subsection{Current measurements (high frequency)}
\begin{figure}[t]
  \centering
  \includegraphics[width=0.4\textwidth]{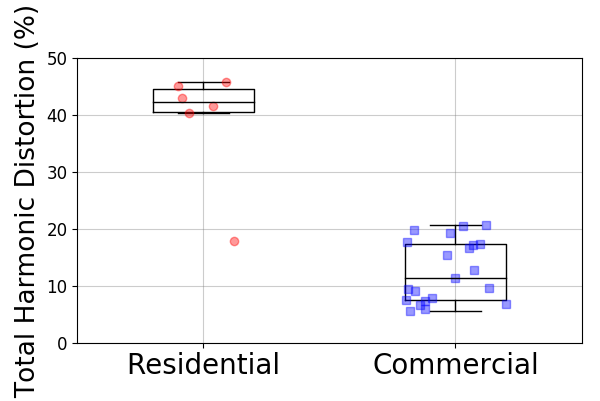}
  \caption{\label{<thd>} Total Harmonic Distortion of current signals for all the "current" datasets (see data column in Table \ref{<datasets>})}
\end{figure}

In buildings the voltage can be considered as pure sine wave. In frequency domain this is characterized by a signal with energy only on the fundamental frequency and no energy on harmonic frequencies. On the contrary, the current signal shows relatively important energies on harmonic frequencies due to non linear devices present on the network. This property can be measured with the total harmonic distortion (THD). It is based on the coefficients of the discrete Fourier transform (DFT) of the current signal. The DFT and the THD are computed for every period:
\begin{align}
\text{THD}(t) = 100 \times \frac{\sqrt{\sum_{h=2}^N \currDFT(h,t)^2}}{\sqrt{\sum_{h=1}^N \currDFT(h,t)^2}},
\end{align}
where $\currDFT(h,t)$ is the $h^{th}$ coefficient of the DFT of $\curr(.,t)$.
Figure \ref{<thd>} shows lower values for commercial buildings that may be explained by an important proportion of linear induction motors (heating, ventilation or air conditioning) which do not have harmonics energy.

\section{A High Frequency Current Waveforms Model}
\label{sec:model}
In this section, we develop a physically-inspired data model that will enable us to reproduce the behaviour of the electrical network of a building in a bottom up procedure. It is based on a building model, category models and individual devices models.

\subsection{The building model}
\label{sub:building_model}
The model that we put forward in this section relies on several hypotheses. 
First, all the electrical devices are supposed to be plugged in parallel on the network: the current waves observed on the root node of the network are then the sum of the currents of all devices. This is a direct application of the Kirchoff's current law.
Then, the electrical network is supposed to be in ideal conditions: the voltage is considered to be identical on each node of the network and independent from the current. Moreover, in the following, all current signals of devices are supposed to be independent. This assumption holds only if the voltage signal is purely periodic since the current waveform depends on the voltage waveform for most devices: $\forall t, \ \volt \left(n, t\right)=\volt_0 \left(n \right)$.

Finally, for the sake of simplicity, only single-phase electrical networks are considered here, but three-phase networks can be simulated in a similar fashion.

These assumptions lead us to the following model for total current: 
\begin{equation}  \label{eq:building_model}
\curr \left( n, t \right) = \sum_{c \in \mathrm{C}} \curr_c \left( n, t \right) + \epsilon \left( n, t \right)
\end{equation}
where $\curr$ is the total current measured at the root node of the network, $\curr_c$ is the current signal of a category $c$ of appliances, $\mathrm{C}$ is the ensemble of category indices, and $\epsilon(n, t)$ is a zero-mean Gaussian noise.
 
\subsection{The category model}
\label{sub:category_model}
Since the number of identical equipments can be high in large buildings (\textit{e.g.} corridors light bulb, computers or resistive heaters), it is often more important to evaluate a whole category consumption instead of each single device consumption (especially for specific NILM applications such as energy management). We then define herein a category as the aggregation of one to many similar equipments as follows:
\begin{equation} \label{eq:cat_model}
\curr_c \left( n, t \right) = \sum_{d \in \mathrm{D}_c} \curr_{c,d} \left( n, t \right)
\end{equation}
where $\curr_{c, d}$ is the current of device $d$ belonging to category $c$. $\mathrm{D}_c$ corresponds to the set of devices belonging to category $c$.  

\subsection{The device model}
\label{sub:device_model}
Finally, the current of a particular device is modeled using a factorization-based approach, given as follows: 
\begin{equation} \label{eq:dev_model_simple}
\curr_{c,d} \left(n, t \right) = \sig_{c,d} \left(n\right) \act_{c,d} \left(t \right)
\end{equation}
where $\sig_{c,d}$ and  $\act_{c,d}$ are respectively called the \textit{current waveform signature} and the \textit{activation} of device $d$ of category $c$. 
The waveform signature corresponds to a fixed pattern that describes the typical current response to the voltage. 
The activation is a nonnegative magnitude and its nature depends on the type of devices ($0$ / $1$ function or continuously varying). 
As it will be demonstrated in Section \ref{sub:dev_mod_eval}, we extend our model for more complex devices by enabling the use of more than one signature in the factorization:
\begin{equation} \label{eq:dev_model}
\curr_{c,d} \left(n, t \right) = \sum_{k=1}^{K_{c,d}} \sig_{c,d} \left(n, k \right) \act_{c,d} \left(k, t \right)
\end{equation}
$K_{c,d}$ is the number of signatures and activations used to model device $d$. The number of components used in the model ($K_{c,d}$) and the nature of the activations ($\act_{c,d}$) enable us to classify the devices into 4 main classes, as shown in Table \ref{<devices>}. In the literature \cite{hart1992,klemenjak2016non}, the common devices' taxonomy includes only 3 classes: (i) on/off or constant device, (ii) multi-state and (iii) continuously varying. This approach is based on low frequency features of load curves whereas we take high frequency characteristics into account. We can see in Table \ref{<devices>}, that the main difference is that the original continuously varying class has been divided into two classes depending on the number of signatures used to model it (see Section \ref{sub:dev_mod_eval}).
\begin{table}[t]
  \centering 
  \caption{A new devices' taxonomy based on high frequency current features.}
  \label{<devices>}
  \begin{tabular}{|c|c|c|}
      \hline
      Activations & \multirow{2}{*}{Simple} & \multirow{2}{*}{Complex} \\
      Signatures & & \\
      \hline
      Unique & On/Off or Constant & Varying load \\
      \hline
      Multiple & Multi-state & Varying signature \\
      \hline
  \end{tabular}
\end{table}

To fix the inherent ambiguity of the multiplicative model of \eqref{eq:dev_model}, we normalize the signatures such that: 
\begin{equation}
\forall\: c,d,k\quad\frac{1}{N} \sum_{n=1}^{N}\sig_{c,d}(n,k).\volt_0(n)=1 
\end{equation}
It has the double advantage to fix the multiplicative ambiguity and to directly link the activations to the consumed power. Indeed, the power consumption of device $d$ is given by: 
\begin{equation}
\label{eq:norm}
\begin{aligned}
\pwr_{c,d}(t) &= \frac{1}{N}\sum_{n=1}^N \curr_{c,d}(n,t).\volt_0(n) \\
& = \sum_{k=1}^{K_{c,d}} \act_{c,d}(k,t)\frac{1}{N}\sum_{n=1}^N \sig_{c,d}(n,k).\volt_0(n) \\
& = \sum_{k=1}^{K_{c,d}} \act_{c,d}(k,t)
\end{aligned}
\end{equation}
We can notice that in the case of a device with a single component ($K_{c,d}=1$), the activation becomes the power consumption. Otherwise, the power equals the sum of the activations of each component.

\subsection{The overall model}
\label{sub:generative_process}
Combining the individual models \eqref{eq:building_model}, \eqref{eq:cat_model} and \eqref{eq:dev_model} gives the model for the total current:
\begin{equation}
\curr \left(n, t \right) = \sum_{c \in \mathrm{C}} \sum_{d \in \mathrm{D}_c} \sum_{k=1}^{K_{c,d}} \sig_{c,d} \left( n, k \right)\act_{c,d} \left(k, t \right) + \epsilon \left( n, t \right).
\end{equation}
Finally, taking into account equations \eqref{eq:power_def} and \eqref{eq:norm}, we obtain the following formula for the power per category:
\begin{equation}
\pwr_c \left( t \right) = \sum_{d \in \mathrm{D}_c} \sum_{k=1}^{K_{c,d}} \act_{c,d}\left(k,t\right). \label{eqn:gen_model}
\end{equation}

\subsection{Device model evaluation}
\label{sub:dev_mod_eval}
We have seen that our model for current waveforms is based on real signatures and on nonnegative activations. In this context, we propose to apply the semi non-negative matrix factorization (SNMF) algorithm developed by Ding et al. \cite{ding2010convex} to estimate signatures and activations from individual equipment measurements and then evaluate the goodness of fit of our model.

The SNMF model applied to NILM settings consists in solving the following optimization problem \cite{ding2010convex}:
\begin{align}
\min_{S, A}\| I - SA^T \|^2_{\text{Fro}} , \qquad \text{such that } A \geq 0
\end{align}
where $I \in \mathds{R}^{N \times T}$ is the current observation matrix of a device $d$ in category $c$ (we have dropped the device and category subscripts for the sake of clarity), $S \in \mathds{R}^{N \times K}$ is the signature matrix, $A \in \mathds{R}_{+}^{K\times T}$ is the activation matrix, and $\|M\|_{\text{Fro}}$ denotes the Frobenius norm.

We now define the Signal to Noise Ratio (SNR) which is a metric for measuring the goodness of fit of our model on real datasets:
\begin{equation}
 \text{SNR} = 10\times\log_{10}{\left(\frac{\sum_{n,t}\left(\hat{\curr}(n,t)\right) ^2}{\sum_{n,t}{\left(\curr \left(n, t \right) - \hat{\curr}(n,t)\right) ^2}}\right)},
\end{equation}

where $\hat{\curr}(n,t)=\sum_{k=1}^{K}\hat{\sig} \left(n, k \right) \hat{\act} \left(k, t \right)$ is the model reconstruction using the estimated signatures ($\hat{\sig}$) and activations ($\hat{\act}$).

The efficiency of our device model is evaluated using two high frequency public datasets which correspond to a few seconds of high frequency current measurements following the switching ON of a device \cite{gao2014plaid,picon2016cooll} and a third private dataset which corresponds to high frequency current measurements over several days for devices in commercial buildings.

\begin{figure}[t]
  \centering
  \includegraphics[width=0.9\textwidth]{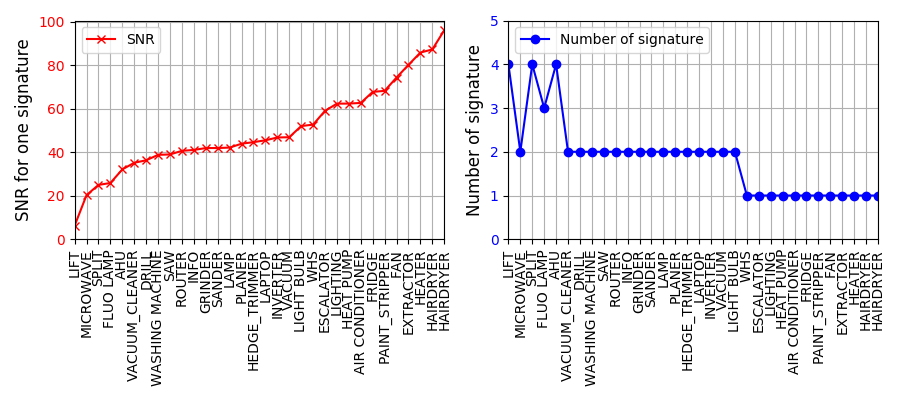}
  \caption{\label{<fact_mod_eval>} Device model evaluation: (left) shows the reconstruction SNR with one signature and (right) shows the minimum number of signatures to use for reaching a SNR of at least $50$ dB.}
\end{figure}

\begin{figure*}[h]
  \centering
  \subfigure["on/off"]{\includegraphics[width=0.41\textwidth]{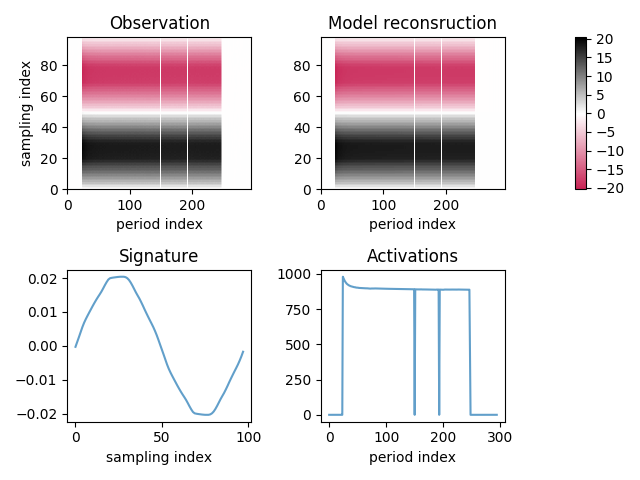}}
  \subfigure[multi-state]{\includegraphics[width=0.41\textwidth]{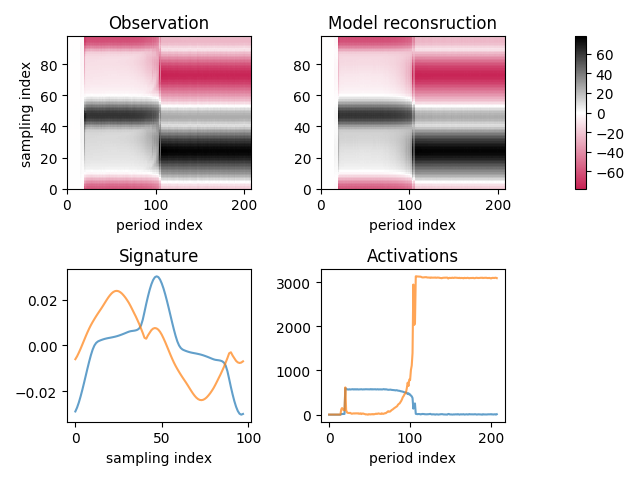}}
  \subfigure[varying load]{\includegraphics[width=0.41\textwidth]{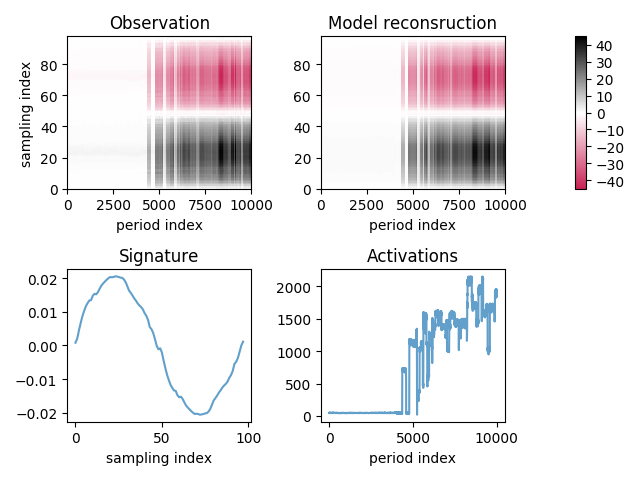}}
  \subfigure[varying signature]{\includegraphics[width=0.41\textwidth]{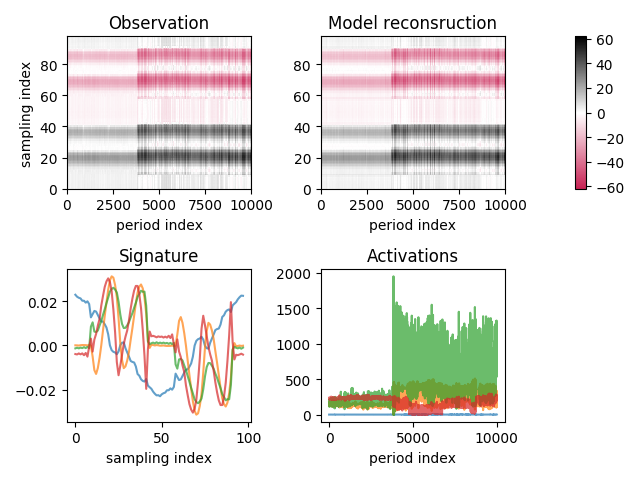}}
  \caption{Learned factorizations for the 4 device classes, each of them is composed of (top left) the observations in matrix shape (sampling index $\times$ period index), (top right) the model reconstruction (sampling index $\times$ period index), (bottom left) the signature columns and (bottom right) the activation rows. The number of signatures is selected such that the SNR $> 50$ dB.}
  \label{<facto_cat>}
\end{figure*}

Figure \ref{<fact_mod_eval>} shows that for several device categories only one component ($K=1$) results in very high values of the SNR, which means a good data reconstruction. For more complex devices, Figure \ref{<fact_mod_eval>} also shows the required number of components to reach a SNR value of at least $50$ dB (excellent reconstruction). Figure \ref{<facto_cat>} illustrates the factorizations learned on the four kinds of devices as presented in Table \ref{<devices>}. This experiment demonstrates the capacity of our model to catch the features of real current measurements. It also validates our choice to separate simple ($K=1$) from complex ($K>1$) devices in our models. It can be noticed that complex devices are for the majority found in commercial buildings and not in residential ones (e.g. air handling unit, lift, split, inverter).

\section{A Generative Procedure for Dataset Simulations}

\newcommand\mycommfont[1]{\footnotesize\ttfamily\textcolor{blue}{#1}}
\SetCommentSty{mycommfont}

In order to be able to simulate new datasets, we need to solve two more problems. First of all, the SNMF model used to estimate factors (signatures and activations) is analytical and do not provide any generating procedure to simulate new data. Secondly, the lack of publicly available high frequency datasets of individual equipments makes it difficult to learn both signature and activations on the same dataset. To circumvent these issues, we first propose separate generative models for signatures and activations. Then, we estimate their parameters and simulate new data independently for signatures and activations using different datasets: (i) short high frequency current measurements for signatures \cite{gao2014plaid,picon2016cooll} and (ii) long low frequency power measurements for activations (\cite{reinhardt2012accuracy} and our private dataset).

\subsection{Signature Sampling Algorithm}
As for the signatures, we use a Gaussian distribution whose mean is given by the templates learned on high frequency current datasets as it is done in Section \ref{sub:dev_mod_eval} with the SNMF algorithm:
\begin{equation}
\sig^{new}(n,k) \sim \mathcal{N}(\hat{\sig}(n,k), \sigma ^2)
\end{equation}
where $\hat{\sig}(n,k)$ is the template learned and $\sigma$ is an hyperparameter.
Figure \ref{<facto_cat>} shows four examples of learned signatures.

\subsection{Activation Sampling Algorithm}
We describe here two different algorithms to simulate the activations: one for simple activations (on/off) and one for complex activations (continuously varying devices).

\subsubsection{Simple activations}
As mentioned in Section 2, a key feature of the activations is their temporal structure. Dinesh et al. \cite{dinesh2017incorporating} introduced a time-of-day usage pattern for a device defined by the probability of being activated at different periods of the day. These `periods of the day' are defined as subsets of a partition of the time. In this study, we follow a similar procedure and partition the time in hours. For instance, one subset (a \textit{period of the day} noted $\mathcal{S}_\tau$) may correspond to the slot 10 am to 11 am for every day. The total number of subsets is hence 24.
We extend Dinesh's approach by providing a generative model for on/off device activations. We are considering here $0$ or $1$ activations and use a deterministic switching mode 2-state Markov chain to model the device's activation where the transition probability is defined as: 
\begin{equation}
\forall \tau, \forall t \in \mathcal{S}_\tau, \forall i,j \in \{0, 1\}^2 \quad P[\act(t)=i|\act(t-1)=j]=\gamma_\tau(i,j),
\end{equation}
where $t$ is the time index, $\mathcal{S}_\tau$ is a period of the day and $\gamma_\tau$ the transition matrix for period of the day $\mathcal{S}_\tau$. This model enables us, first, to infer the transition probabilities depending on the period of the day from databases and, second, to generate new activations.
Using maximum likelihood inference, the $\gamma$ parameter is estimated by the following equation:
\begin{equation}\label{eq:inference}
  \hat{\gamma}_\tau(i,j)=\frac{\sum_{t\in \mathcal{S}_\tau}\mathbbm{1}_{(\act(t)=i, \act(t-1)=j)}}{\# \mathcal{S}_\tau},
\end{equation}
where $\# \mathcal{S}_\tau$ is the size of subset $\mathcal{S}_\tau$.
Intuitively, this estimation corresponds to counting the number of ON-to-OFF and OFF-to-ON events occurring during the period of the day $\mathcal{S}_\tau$.
We are using a public dataset gathering power measurements for individual devices for several days to estimate the parameters \cite{reinhardt2012accuracy}. 
Firstly, we transform the power time series into on/off time series using a simple thresholding mechanism: $\tilde{x}(t)=\mathbbm{1}_{(x(t) > 20)}$. Secondly, we estimate the model parameters using \eqref{eq:inference}. Finally, the learned parameters are used to generate new data:
\begin{align}
 \act^{new}(t) \sim \mathcal{B}\text{er}( \hat{\gamma}_\tau(1,\act^{new}(t-1))) 
\end{align}
where $\mathcal{B}\text{er}$ is the Bernoulli distribution.

\begin{figure*}[t]
  \centering
  \subfigure[TV-LCD]{\includegraphics[width=0.49\textwidth]{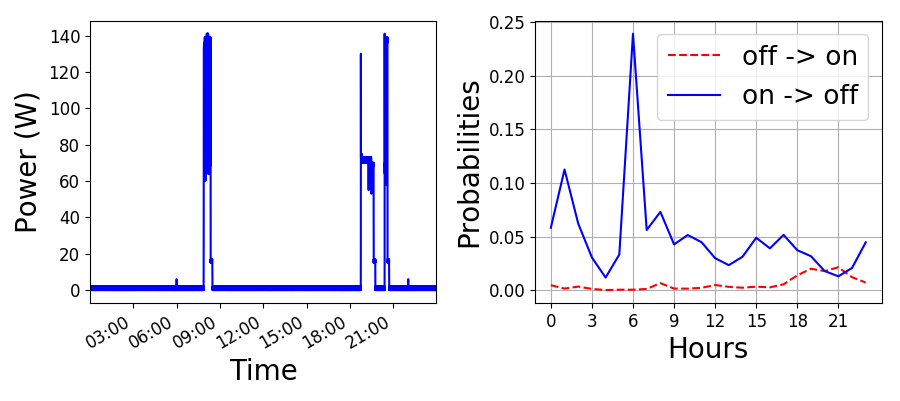}} 
  \subfigure[Coffee maker]{\includegraphics[width=0.49\textwidth]{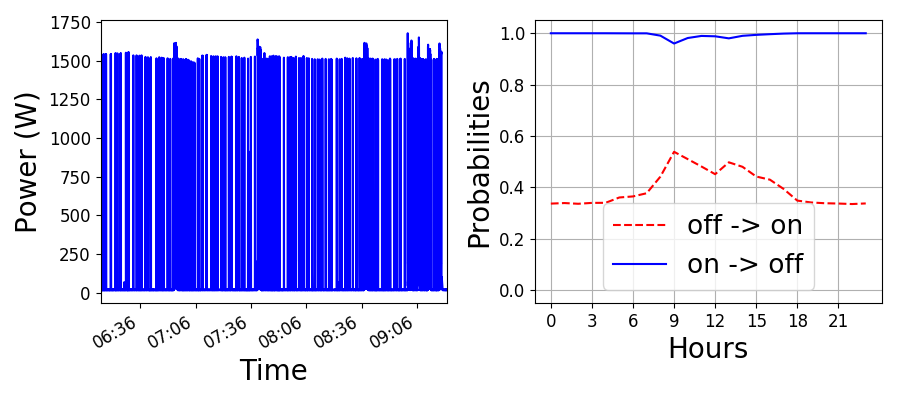}}
  \caption{Activation probabilities learned on public dataset (right) and a few hours of the measurements (left).}

  \label{<simple_activations>}
\end{figure*}

Figure \ref{<simple_activations>} shows two examples of simple activations data and the learned activations parameters. The learned activations show that the probability of switching ON is highest at 8 am and 7 pm for the TV. It also shows that for the coffee maker, the probability of switching ON is quite high all day long and that once ON it immediately switches OFF.

\newcommand{\tmpsize}{0.24}

\subsubsection{Complex activations}
\begin{figure*}[t]
  \centering
  \subfigure[IT devices]{\includegraphics[width=0.49\textwidth]{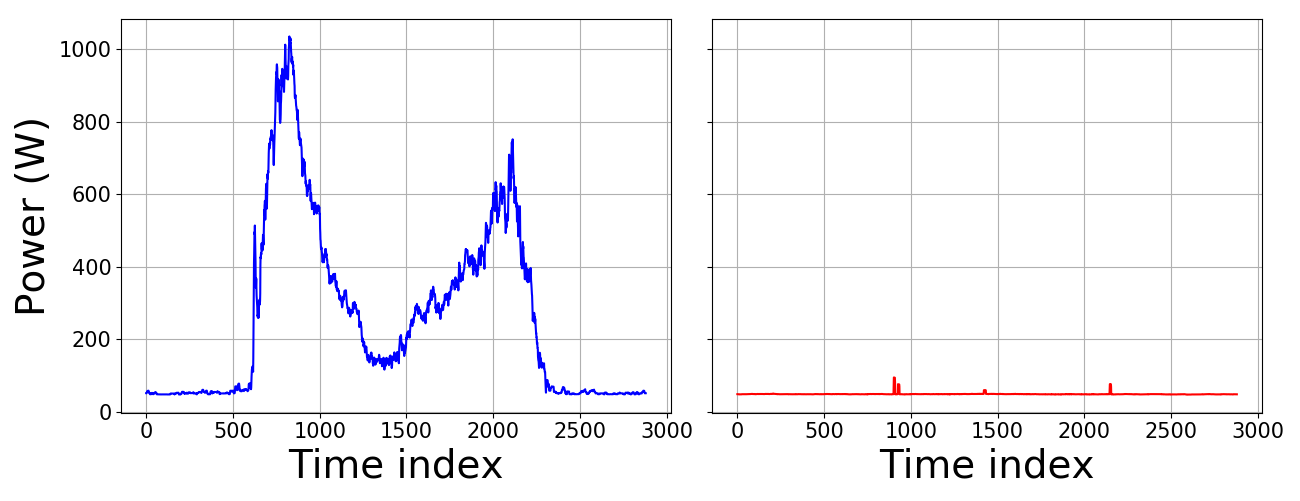}}
  \subfigure[Heat pump]{\includegraphics[width=0.49\textwidth]{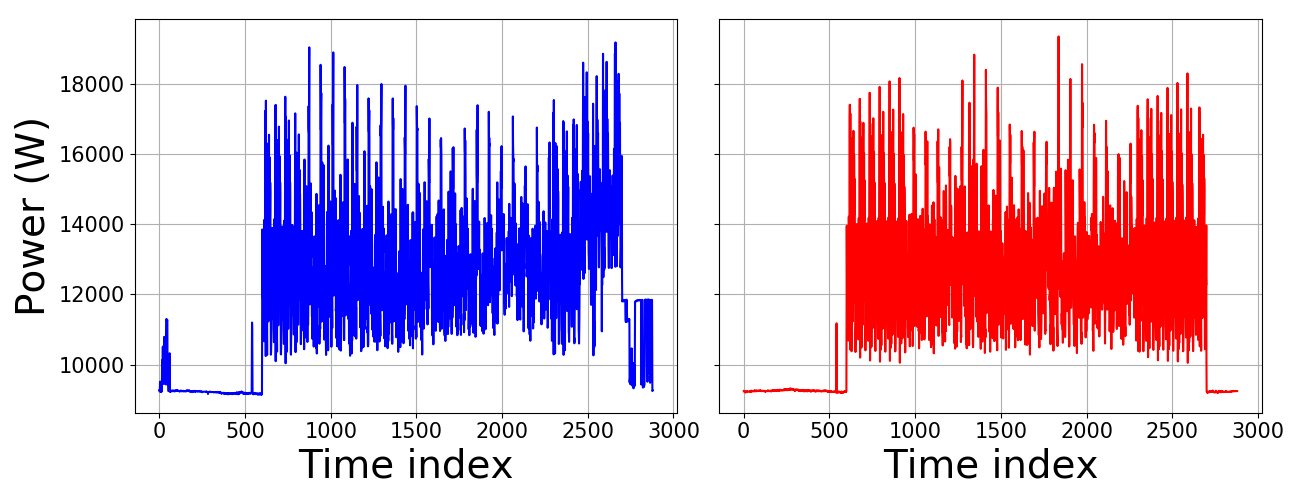}}
  \caption{Activation templates learned on the private dataset, the templates correspond to one day (timestep = 30 sec): week-day (left) and day-off (right).}
  \label{<templates>}
\end{figure*}

In this part, we are considering generating activations by learning `activation templates' on a private dataset due to the lack of public dataset for this kind of devices. The private data is collected from two large commercial buildings in two different cities in France. It contains $11$ device categories and is recorded during several weeks at low sampling frequency. The goal of the templates is to catch the typical power consumption of a device category during a \textit{period of the day} and thus account for the daily seasonal effects shown in commercial buildings (see Section \ref{sec_analysis}). Since many equipments are programmed to switch on or off on particular days (air handling unit, heaters) or depend on building occupancy (computers), we distinguish the week days and the days off. In this part the partition of the time is made with period of $30$ seconds. The total number of subsets is then $5760$ ($2880$ periods of $30$ seconds per week days and days off). In order to compute such templates, we simply average the power consumptions of individual devices over several weeks of data per period of the day:
\begin{equation}
  \hat{\act}(\tau) = \frac{\sum_{t \in \mathcal{S}_\tau}{\pwr(t)}}{\# \mathcal{S}_\tau}
\end{equation}
The learned templates are illustrated in Figure \ref{<templates>}. We can observe that IT devices are switched off during day off and have smooth load curves whereas the heat pump has a more noisy consumption.

To generate new data, we multiply a positive noise with the templates to take the day to day variability into account:
\begin{equation}
  \forall \tau, \forall t \in \mathcal{S}_\tau \quad \act^{new}(t) = \hat{\act}(\tau) \times \exp(\epsilon(t)),
\end{equation}
where $\epsilon$ follows an ARMA(p,q) process \cite{box2015time}. An ARMA process is a linear time serie model involving lagged values of itself and of a white noise. It is well used in time series modeling because of its stationary property and its ability to model autocorrelation at different lags. We used it to add smooth variations to the templates from one day to another.

In Section \ref{sub:device_model}, we defined two kinds of devices with complex activations: (i) single signature or (ii) multiple signatures. While the former has just been addressed, we need to find a generative process for the latter. The proposed generative method uses the same process as before and considers a random convex combination of the activations:
\begin{equation}
  \forall \tau, \forall t \in \mathcal{S}_\tau \quad \act^{new}(k,t) = \hat{\act}(\tau) \times \exp(\epsilon(t)) \times \delta(k), 
\end{equation}
where $\delta$ is a K-dimensional Dirichlet-distributed random variable whose parameter $\alpha = (\alpha_1,\dotsc,\alpha_K)$ controlls the activation components proportion and $\alpha$ is considered as an hyperparameter. Note that $ \sum_k \delta(k) = 1$ and $\delta(k) \geq 0$ for all $k$.

\section{A Synthetic High-frequency Energy Disaggregation (SHED) dataset for commercial buildings}

In order to enable high frequency NILM algorithm evaluation, we release a synthetic dataset called SHED. The purpose of our simulations is to evaluate the disaggregation performance of NILM algorithms (i.e. the capability to separate individual consumptions from a mixture). Our simulation procedure does not allow the evaluation of classification performance (assigning every disaggregated load curve to a particular category).

\begin{table}[t]
  \centering
  \caption{\label{<shed>} Devices used to simulate the buildings in the SHED dataset: On/Off (A), Multi-state (B), Varying load (C), Varying signature (D).}
  \begin{tabular}{|l|c|c|c|c|c|}
      \hline
      Class & A & B & C & D & Total \\
      \hline
        building 1&4&0&2&3&9\\
        building 2&1&4&2&3&10\\
        building 3&0&2&2&3&7\\
        building 4&2&0&4&3&9\\
        building 5&0&3&4&1&8\\
        building 6&3&0&3&4&10\\
        building 7&0&0&3&2&5\\
        building 8&0&0&4&4&8\\
      \hline

  \end{tabular}
\end{table}

The SHED dataset consists of 8 buildings. For each building, it includes the total current consumption, as well as the individual consumptions corresponding to different categories. For buildings 1 to 6, the individual consumptions consist of low frequency power measurements and for buildings 7 and 8 they consist of high frequency current measurements. One current waveform is recorded at every $30$ seconds and for every current waveform $200$ points are sampled. Power measurements are also sampled at $1/30 Hz$. The choice of the classes of the devices and the number of categories enables us to control the complexity of each building: the details of the buildings are described in Table \ref{<shed>}. The features of the buildings have been selected in such a way that they would correspond to commercial buildings.

\begin{figure*}[!]
  \centering
  \subfigure[Autocorrelation]{\includegraphics[width=0.4\textwidth]{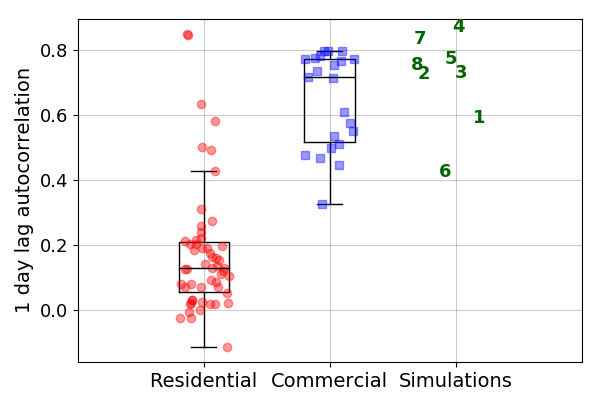}} 
  \subfigure[Kurtosis]{\includegraphics[width=0.4\textwidth]{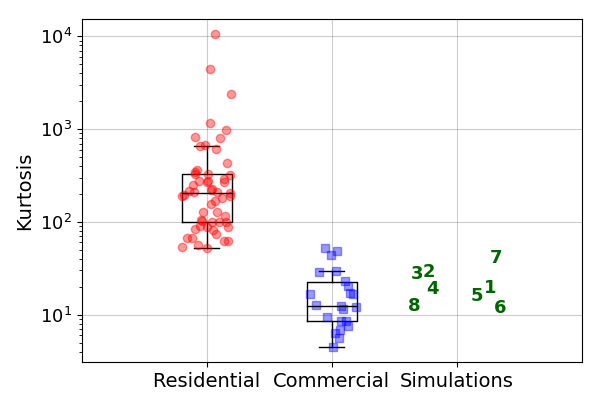}}
  \subfigure[Entropy]{\includegraphics[width=0.4\textwidth]{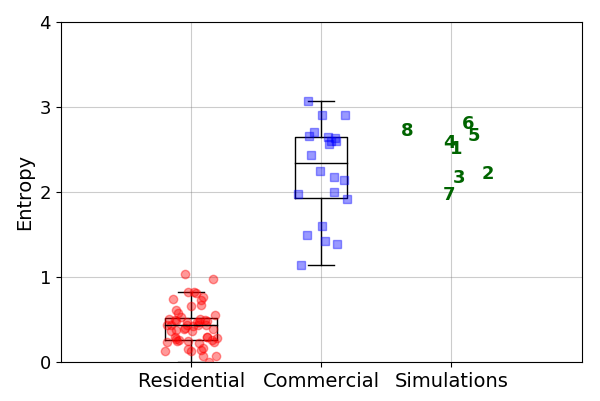}}
  \subfigure[Laplace scale parameter]{\includegraphics[width=0.4\textwidth]{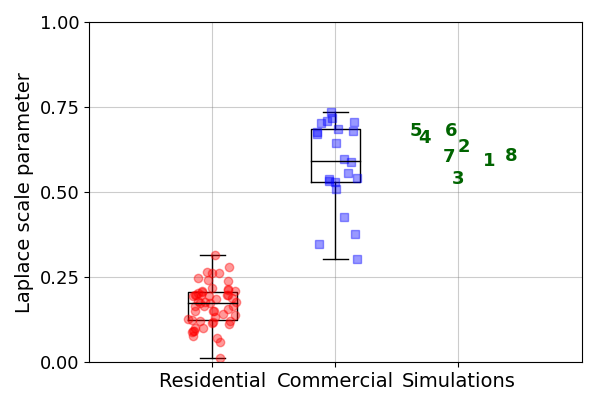}}
  \subfigure[THD]{\includegraphics[width=0.4\textwidth]{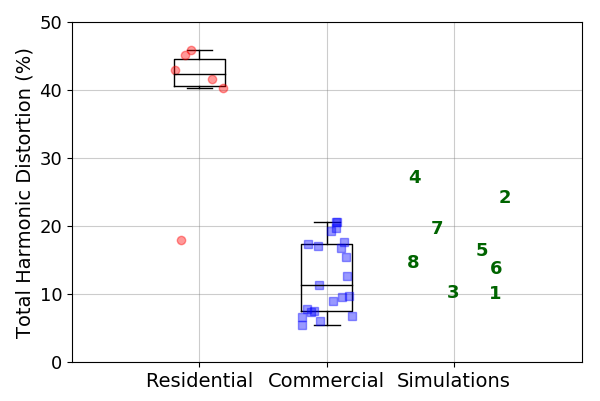}}
  \caption{Quantitative evaluation of the simulated datasets: comparison of the statistical metrics of simulations and real datasets. Every circle or square corresponds to one building. Numbers in simulations columns correspond to building indexes in the SHED dataset.}
  \label{<db_eval>}
\end{figure*}

After having evaluated the quality of the device model in Section \ref{sub:dev_mod_eval}, we evaluate here the total current of the building. We use the metrics introduced in Section \ref{sec_analysis} to check the quality of the simulations. Figure \ref{<db_eval>} shows clearly that the simulated datasets share very similar statistical properties as real commercial datasets. It provides a strong justification that our simulations are realistic. We can however notice that the THD values of simulations are more spread than for commercial buildings. It may be explained by the fact that the public datasets used for simulating signatures mostly correspond to residential equipments.

Finally, the SHED dataset can be downloaded at \website.

\section{Conclusion and Discussion}

We addressed the task of non-intrusive load monitoring for commercial buildings by analyzing existing datasets, developing a synthetic data generation process, and releasing an evaluation dataset.

We produced an extensive data analysis on public and private datasets that showed that commercial and residential buildings have significantly different characteristics. The study of the power derivative distribution illustrated that the residential distributions are more peaky at zero than the commercial ones. On top of this, we showed that the kurtosis, the entropy and the Laplace scale parameter of the power derivative are good discriminative indicators for residential and commercial buildings. We explained this difference by a higher amount of devices in commercial buildings and the presence of complex categories of devices (continuously varying equipment, multitude of similar devices). These statistical characteristics are in contradiction with the hypothesis used for residential NILM algorithms (`one at a time' and `constant load'). In this context, detecting a single event on the power curve is a difficult task and this explains why residential NILM algorithms fail when applied to commercial buildings. The statistical metrics used in our study suggest that using a soft version of the "one at a time" hypothesis such as "few at a time" (only a few devices are responsible for the power variations at every instant) would be more realistic.

Motivated by the lack of data for commercial buildings, we developed a generative model for synthesizing high frequency current waveforms. Inspired by physical realities, it is compound of three layers: devices, categories and buildings. Our device model is based on a matrix factorization approach, breaking down high frequency current waveforms into signatures and activations components. The model efficiency has been validated with real data. It also enabled us to introduce a new device taxonomy taking the high frequency features of the devices into account.

Finally, we proposed a simulation procedure that enables us to learn parameters on real data and then simulate new synthetic data. Our quantitative evaluation experiments showed that the simulated datasets share the same statistical properties as real datasets. To enable algorithms testing and comparison, a simulated dataset called SHED is released at \website.
\section*{References}
\bibliographystyle{elsarticle-num}
\bibliography{nilm} 

\end{document}